TITLE: Interfertile oaks in an island environment: I. High nuclear genetic differentiation and high degree of chloroplast DNA sharing between Q. alnifolia and Q. coccifera in Cyprus. A multipopulation study.

AUTHORS: Charalambos Neophytou*[1,2], Aikaterini Dounavi[1], Siegfried Fink[2], Filippos A. Aravanopoulos[3]

CORRESPONDING AUTHOR:
Charalambos Neophytou
Forest Research Institute of Baden-Württemberg
Department of Forest Ecology
Wonnhaldestr. 4
D-79100 Freiburg
Germany

E-mail address: Charalambos.Neophytou@forst.bwl.de
(current e-mail address: chneophytou@gmail.com)
Telephone: +49 / 761 / 40 18 159
Fax: +49 / 761 / 40 18 133

---

[1] Department of Forest Ecology, Forest Research Institute of Baden-Württemberg, Wonnhaldestr. 4, D-79100, Freiburg, Germany
[2] Chair of Forest Botany, Faculty of Forest and Environmental Sciences, Albert-Ludwigs University of Freiburg, Bertoldstr. 17, 79085, Freiburg, Germany
[3] Laboratory of Forest Genetics and Tree Breeding, Faculty of Forestry and Natural Environment, Aristotle University of Thessaloniki, P.O. Box 238, Thessaloniki, Greece

ABSTRACT

The evergreen *Quercus alnifolia* and *Q. coccifera* form the only interfertile pair of oak species growing in Cyprus. Hybridization between the two species has already been observed and studied morphologically. However, little evidence exists about the extent of genetic introgression. In the present study, we aimed to study the effects of introgressive hybridization mutually on both chloroplast and nuclear genomes. We sampled both pure and mixed populations of *Q. alnifolia* and *Q. coccifera* from several locations across their distribution area in Cyprus. We analyzed the genetic variation within and between species by conducting Analysis of Molecular Variance (AMOVA) based on nuclear microsatellites. Population genetic structure and levels of admixture were studied by means of a Bayesian analysis (STRUCTURE simulation analysis). Chloroplast DNA microsatellites were used for a spatial analysis of genetic barriers. The main part of the nuclear genetic variation was explained by partition into species groups. High interspecific differentiation and low admixture of nuclear genomes, both in pure and mixed populations, support limited genetic introgression between *Q. alnifolia* and *Q. coccifera* in Cyprus. On the contrary, chloroplast DNA haplotypes were shared between the species and were locally structured suggesting cytoplasmic introgression. Occasional hybridization events followed by backcrossings with both parental species might lead to this pattern of genetic differentiation.

# INTRODUCTION

Since DNA markers became widely available, genetic research of natural oak populations has been intensified. A rapidly growing number of studies describing molecular genetic diversity and differentiation have been carried out (Muir et al. 2000; Craft and Ashley 2006; Soto et al. 2007). A wide variety of both nuclear and organelle DNA markers have been applied, leading to contrasting results. On one hand, nuclear markers have been successfully used to differentiate at the species level (Muir et al. 2000; Curtu et al. 2007a). On the other hand, chloroplast DNA lineages have been found to be shared between related species (Petit et al. 2002; Lumaret et al. 2002; López de Heredia et al. 2005). The underlying evolutionary forces leading to this pattern have been intensively debated. Both common ancestry and interspecific gene flow have been proposed as contributing to the high degree of genetic information sharing (Muir and Schlötterer 2005; Lexer et al. 2006). Increasing evidence supports that hybridization plays a central role in the genomic architecture of the different taxa not only in oak, but also in several other plant genera (Kremer et al. 2002; Petit et al. 2004; Nosil et al. 2009).

The genus *Quercus* has served as a model for studying hybridization since the advent of evolutionary thinking (Darwin 1859). Many oak species that belong to the same phylogenetic group are interfertile and hybridize under natural conditions. Despite high levels of interspecific gene flow, intermediate individuals often occur in very limited numbers. For instance, Rubio de Casas et al. (2007) found that *Q. coccifera* maintains its morphological identity despite high levels of hybridization with *Q. ilex* in Spain. On the other hand, large scale studies based on chloroplast DNA markers reveal high levels of cytoplasmic sharing between related species. An extensive study including 2600 populations of European oak species of the subgenus *Quercus* s.s. showed that chloroplast DNA haplotypes were only geographically structured and no correlation with species was found (Petit et al. 2002). Similarly, chlorotypes in Mediterranean sclerophyllous oak species present a geographic distribution along an east-west axis and are shared among species (Lumaret et al. 2002; López de Heredia et al. 2007).

Common structures of organelle DNA genomes support past hybridization and introgression events (Lumaret and Jabbour-Zahab 2009). Especially in the well studied European oak species of the subgenus *Quercus* s.s., maternal lineages carry the imprints of the postglacial recolonization and are entirely shared between species. No correlation between nuclear and chloroplast DNA variation was found, which supports the notion that species regain their identity after a limited number of backcrossings following the initial hybridization event (Kremer et al. 2002; Petit et al. 2004). Similar cases of 'cytoplasmic capture' have been described between the occasionally hybridizing Mediterranean species *Q. ilex* and *Q. suber*. In this case, it was shown that some populations of *Q. suber* possess chloroplast DNA lineages that originated from sympatric or neighbouring *Q. ilex* populations (Belahbib et al. 2001).

Furthermore, it has been demonstrated that interspecific differentiation between related oak species varies from locus to locus (Curtu et al. 2007a; Soto et al. 2007; Gugerli et al. 2008). This differentiation pattern was attributed either to past selective sweeps (Muir and Schlötterer 2005) or to active selection at adaptive loci following hybridization (Scotti-Saintagne et al. 2004). The latter process leads to a fast 'resurrection' of the parental species after an initial hybridization event (Petit et al. 2004). Rubio de Casas et al. (2007) in a case study of *Q. coccifera* and *Q. ilex* support that parental phenotypical traits re-emerge in early stages of introgression in a similar way, whereas at selectively neutral loci a large amount of genetic variation remains shared.

In Cyprus, *Q. coccifera* and the endemic *Quercus alnifolia* are significant elements of the local Mediterranean sclerophyllous flora. They are the only two interfertile oak species and do not hybridize with *Q. infectoria*, the third indigenous oak species of the island (Neophytou et al. 2008). *Quercus alnifolia* is abundant in the central mountainous area of Troodos and grows on volcanic geological substrate. *Quercus coccifera* has a wider, but more scattered distribution and is locally dominant on calcareous formations (Barbero and Quezel 1979). However, the two species grow over large areas in sympatry and overlapping of their flowering periods allows interspecific pollination. Hybridization between the two species has been studied by morphological traits only (Knopf 2006, Hand 2006, Neophytou et al. 2007). Morphologically intermediate individuals were rather scarce. A multivariate analysis using morphological leaf traits supports some interspecific introgression in sympatric populations and reveals a higher degree of diversity in *Q. coccifera* (Neophytou et al. 2007). Additionally, according to a study describing variation at seven isoenzyme loci, *Q. alnifolia* shares all its alleles with *Q. coccifera* and exhibits markedly lower diversity at most loci (Toumi and Lumaret 2001).

In our study, genetic variation within and between *Q. alnifolia* and *Q. coccifera* was investigated using both chloroplast and nuclear DNA microsatellite markers. For this purpose, pure and sympatric populations were included covering a large part of the species' distribution area in Cyprus. We studied the effects of hybridization and subsequent genetic introgression on their nuclear and chloroplast genomes. We addressed following questions:
- Is there substantial genetic differentiation between the two species despite hybridization? Is there intraspecific subdivision?
- Does interspecific differentiation vary within mixed and between pure populations?

Is there introgressive hybridization?
- Does cpDNA variation follow a different interspecific and intraspecific variation pattern?

MATERIALS AND METHODS

*Sample collection*

Samplings comprised six pure *Q. alnifolia* populations, six pure *Q. coccifera* populations and one mixed stand with both species. One large sample (96 individuals) was taken from one pure population per species, in order to calculate population diversity measures (localities Stavros for *Q. alnifolia* and Kannaviou for *Q. coccifera*, a1 and c1 in Figure 1 respectively). Additionally, 207 *Q. alnifolia* (a2 in Figure 1), 66 *Q. coccifera* (c2 in Figure 1) and four individuals with intermediate morphological characteristics were collected from a *Q. alnifolia*-dominated mixed stand (locality Kambos) in order to detect possible genetic introgression. Smaller samples (5-12 individuals) were taken from five additional pure populations per species scattered throughout the distribution area of each species in Cyprus, in order to capture as much chloroplast DNA variation as possible (a3-a7 in *Q. alnifolia* and c3-c7 in *Q. coccifera* in Figure 1). Regarding the large populations, subsamples were used for the cpDNA analysis. In particular, 24 samples were analyzed from each one of the pure populations (a1 and c1), whereas 28 *Q. alnifolia* and eight *Q. coccifera* individuals from the mixed stand were also included into the analysis. According to Pons and Petit (1995), a small sample size and a large number of sampled populations are required when using markers with high inter- and low intrapopulation variation, as in the case of chloroplast DNA haplotypes. In total, 564 individuals were sampled. All sampled plants were georeferenced using a GPS. Leaf probes were stored in plastic bags with silica gel at 4°C for a short period until DNA extraction.

*Laboratory procedures*

Leaf material was first freeze-dried and then DNA was extracted using the DNeasy 96 extraction kit (Qiagen). Subsequently, PCR was carried out for the amplification of five nuclear and seven chloroplast DNA (cpDNA) microsatellites (SSRs). Nuclear SSR loci QpZAG9 and 119 were first described in *Q. petraea* (Steinkellner et al. 1997) and QrZAG11, 96 and 112 in *Q. robur* (Kampfer et al. 1998). The aforementioned nuclear loci were chosen after testing several SSR primer pairs from the species mentioned above, most of which showed diverse amplification problems. Thus, primer selection was based on the quality of amplification products and reliability of scoring. The volume of the reaction mixture was 25µl containing 10ng of template DNA. Reaction Buffer "E" was used (Genaxxon, Germany). The latter contained 670 mM Tris/HCl (pH 8.8 at 25°C), 160 mM $(NH_4)_2SO_4$ and 0.1 % Tween 20 and was diluted one to ten in the reaction mix. Concentrations of other PCR chemicals were as follows: 1.5 mM $MgCl_2$, 200 mM of each dNTP, 1 unit Taq DNA Polymerase (Genaxxon, Germany) and 0.2 mM Primer (Biomers, Germany). PCR programs included an initial denaturation step at 95°C lasting 8 min, 10 cycles of 94°C for 15 s, an annealing step at 50°C for loci QpZAG119, QrZAG11 and QrZAG112 or 57°C for loci QpZAG9 and QrZAG96, a denaturation step at 72°C for 15 s and 23 additional cycles with reduced denaturation temperature (89°C) and the remaining parameters as in the first 10 cycles. No final elongation was performed. For the chloroplast microsatellites PCR programs comprised an initial denaturation step at 95°C lasting 5 min and 30 cycles with denaturation (94°C), annealing and denaturation steps (72°C) lasting one minute each. Annealing temperatures were set to 50°C for loci µcd4, µcd7, µdt1, µdt3 and µkk3 and 57°C for locus ccmp2. With regards to the cpDNA SSR loci, primers for ccmp2 were initially developed in *Nicotiana tabacum* (Weising and Gardner 1999), and for µcd4, µcd7, µdt1, µdt3 and µkk3 in *Q. petraea* and *Q. robur* (Deguilloux et al. 2003). Allele scoring was carried out by means of capillary electrophoresis using an ABI PRISM 3100 genetic analyzer (Applied Biosystems). For this purpose, primers were fluorescent labelled and the GeneScan ROX-400 (Applied Biosystems) size standard was used in order to size the PCR-products. Allele sizing was performed by applying the GeneMapper v4.0 software (Applied Biosystems).

*Data Analysis*

First, we calculated population diversity measures based on nuclear microsatellites. Only the two large populations of each species were included in these analyses – one pure and one from the mixed stand. The sample size of the remaining small populations (N ≤ 12) precluded an accurate estimation of nuclear diversity measures (Nei 1978). We computed allele frequencies, observed and expected heterozygosity as well as private allele frequency using the GenAlEx 6.1 software (Peakall and Smouse, 2006). We additionally used the FSTAT software (Goudet 1995) to calculate allelic richness after rarefaction for each population and locus, which is a diversity measure independent of the population size (Petit et al., 1998). Rarefaction size was 65 corresponding to the minimum number of genotyped individuals per population and locus. By using the same software we computed inbreeding coefficients ($F_{IS}$) for each locus and population and tested significance by performing 10000 randomizations of alleles among individuals within samples. Furthermore, we used the software Microchecker (Van Oosterhout et al., 2004) to test the occurrence of null (non-amplifiable) alleles, which is indicated by an overall significant homozygote excess. One thousand randomizations of alleles were applied in order to test if this deviation was significant. Frequencies of such 'null alleles' were calculated and adjusted genotypes were produced according to Van Oosterhout et al. (2006).

In order to dissect the genetic variation into its components, an $F_{ST}$ based analysis of molecular variance (AMOVA; Excoffier et al. 1992) was conducted by using the multilocus approach of the Arlequin 3.1 software (Excoffier et al. 2005). Partitioning of genetic diversity within and between species and populations was investigated by including the large populations in the analysis and defining species as groups. For the calculation of a Euclidean distance matrix and sums of squares, two dummy haplotypes are produced by randomly setting the gametic phase for each individual. Sum of square deviations within or among populations are insensitive to the choice of haplotypic phase in the diploid individuals (Michalakis and Excoffier 1996). The significance of covariance components and fixation indices was tested in both cases with a non-parametric permutation procedure (Excoffier et al. 1992):
- Haplotypes among populations and among groups (i.e. species) were permuted for testing interspecific variation.
- Haplotypes among populations within groups were permuted for testing variation among conspecific populations.
- Populations were permuted between groups in order to test variation due to species.
Additionally, by using the same software, pairwise $F_{ST}$s were computed separately for each locus and, by permuting individual genotypes among populations, their significance was tested. In all cases we performed 10 000 permutations.

In order to detect genetic structures and assign individuals to groups, a model-based Bayesian procedure was implemented using the Structure 2.2 software (Pritchard et al. 2000) was used. Data from all loci and all samples were used for calculating posterior probabilities of membership to assumed K subpopulations or groups, without prior information of their population of origin. One-hundred-thousand burn-in periods and 100000 Markov Chain Monte Carlo simulations were performed assuming admixture and correlated allele frequencies. In order to decide for the most likely K, data were ran by setting K= 1, 2, …, 10. Ten runs were performed for each K. Maximum posterior probability (lnP(D)) and minimum deviation between runs were used to infer ΔK, a statistic based on the rate of change of lnP(D) between successive K values and its variation among runs. ΔK corresponds to the second order rate of change of lnP(D) for a given value of K and more accurately detects the uppermost hierarchical level of structure, providing an accurate estimation of population clustering (Evanno et al. 2005). Proportions of membership to the K assumed groups for each individual and population were then calculated. Two thresholds of membership proportion, P(X|K)= 0.8 and P(X|K)= 0.95, were empirically used to detect admixed genotypes.

Finally, an analysis of cpDNA variation was conducted. Chlorotypes (cpDNA haplotypes) were defined as different combinations of alleles among the scored cpDNA microsatellites. Relationships between chlorotypes were investigated by means of a minimum spanning network, produced using the Arlequin 3.1 software (Excoffier et al. 2005) and plotted using the TCS 1.21 software (Clement et al. 2000). Furthermore, spatial genetic barriers were studied based on cpDNA variation using Slatkin's $R_{ST}$ values (Slatkin 1995). Measures of differentiation taking into account the degree of similarity between haplotypes like Slatkin's $R_{ST}$ have been shown to be more efficient in estimating subdivision when maternally inherited organelle genomes of angiosperms are considered. This is due to the higher relative contribution of mutation vs. drift to population subdivision caused by limited dispersal (Petit et al. 2005). Pairwise $R_{ST}$ distances between populations were calculated by using the software Microsat (Minch et al. 1995). Additionally, 100 pairwise $R_{ST}$ tables were produced by bootstrapping. The Monmonier maximum difference algorithm (Monmonier 1973) was implemented by software Barrier 2.2 (Manni et al. 2004) in order to place a predetermined number of genetic barriers across a geometric network connecting all populations. Genetic barriers were set for each of the bootstrapped pairwise $R_{ST}$ tables in order to test significance. For each pair of neighbouring populations the number of times that their border was included in a given barrier (0-100) was tested. For choosing the optimal number of barriers this process was repeated by setting different number of barriers, from 1 to 10. Since our aim was to subdivide the populations in a limited number of spatial groups potentially corresponding to the two species, we tried to keep the number of barriers as low as possible and their significance as high as possible. The mixed stand was not considered at this stage since the geographic locations of the two species coincide and an illustration of a large scale geographic barrier including both is not possible.

RESULTS

*Nuclear microsatellite population diversity and differentiation*

Results from pure and sympatric populations, as well as from pooled *Q. alnifolia* and *Q. coccifera* populations are presented in Table 1. In general, all analyzed loci were polymorphic as revealed by the population diversity measures. Some differences occurred among loci, with QpZAG9 and QpZAG119 being the most variable in both species. Loci QrZAG96 and 112 showed low to medium diversity (Table 1) arising from high frequencies of certain alleles. Locus QrZAG11 was highly variable in *Q. coccifera* ($H_e$= 0.795-0.816), whereas in *Q. alnifolia* one allele ('251') was almost fixed with a frequency of 89.9 % (Figure 2) leading to a very low genetic diversity ($H_e$= 0.156-0.220). Species private alleles were mostly observed in low frequencies. Allele '141' of locus QrZAG96 was the most frequent private allele in *Q. coccifera* (12.4 %). Some diagnostic alleles for species differentiation could be identified. These alleles were prevalent in one species whereas they occurred in very low frequencies in the other. The most characteristic example was observed at locus QrZAG112 where allele '86' occurred with a frequency of 85.6 % in *Q. coccifera* versus 2 % in *Q. alnifolia*, whilst allele '88' displayed the opposite pattern with frequencies reaching 72 % in *Q. alnifolia* and only 3.6 % in *Q. coccifera* (Figure 3).

Inbreeding coefficients ($F_{IS}$) were constantly significant at locus QpZAG119 (P< 0.001). An overall significant excess of homozygotes supported the occurrence of null (non-amplified) alleles. Estimates of null allele frequency varied between 27.7 and 29.7% in *Q. alnifolia* and between 10.9 and 13.4% in *Q. coccifera* (Table 1). Thus, this locus was removed from further analyses. A significant $F_{IS}$ was observed at locus QrZAG11 (P<0.01) in the *Q. coccifera* population from the mixed stand. Null allele frequency was again significant in this case. Adjusted genotypes produced by using the Microchecker software were used in further analyses. In the remaining loci, neither $F_{IS}$ nor the presence of null alleles was significant. However, $F_{IS}$ value in the case of QrZAG112 in *Q. coccifera* from the mixed population was high ($F_{IS}$= 0.225, P= 0.06).

Multilocus AMOVA revealed that the main part of the overall genetic variation was due to partitioning into species groups (Table 2). In particular, the interspecific component was significant ($F_{CT}$= 0.333, P< 0.001) representing 33.26 % of the total variation. Inbreeding coefficient within populations was found to be low ($F_{IS}$= 0.028) but significant (P< 0.01). Fixation indices due to the partition of the species groups into populations were low ($F_{SC}$= 0.008) but significant (P< 0.01) and accounted for 0.55 % of the total variation (Table 2). Investigation of population specific fixation measures further revealed high interspecific differentiation. All interspecific pairwise $F_{ST}$s were high and significant (Table 3). Loci QpZAG11 and 112 showed the highest values due to the prevalence of different alleles in each species. One shared allele in high frequency at locus QrZAG96 resulted in a low but still significant interspecific $F_{ST}$. Values of intraspecific $F_{ST}$s were low and did not exceed 0.05. However, locus QrZAG11 showed a significant differentiation in *Q. alnifolia* (P< 0.01). Within *Q. coccifera* locus QrZAG112 demonstrated the highest $F_{ST}$ value ($F_{ST}$= 0.0497; Table 3) which was significant at the 0.001 level. Moreover, at loci QpZAG9 and QrZAG11 interspecific $F_{ST}$s for *Q. coccifera* were low, but significant at the 0.05 level.

Two well separated species groups were defined after a Bayesian analysis of genetic structure was employed. Maximum posterior probability (lnP(D)) strongly increased between runs with K= 1 and K= 2. For larger Ks, lnP(D) values only slightly increased and tended to plateau, whereas variation among runs of a given K increased. The largest ΔK, calculated according to Evanno et al. (2005), was maximal for K= 2, giving the highest posterior probability of K. The two inferred clusters corresponded to two species groups. High membership proportions were measured for all populations, varying between 0.947 and 0.986 for *Q. alnifolia* and between 0.940 and 0.986 for *Q. coccifera* (Table 4). In large samples, whether pure or sympatric, proportion of membership to species groups was high measuring for *Q. alnifolia* 0.975 in Stavros (a1) and 0.966 in Kambos (a2) and for *Q. coccifera* 0.967 in Kannaviou (c1) and 0.978 in Kambos (c2). Admixed genotypes (P(X|K)<0.95) were observed in almost all populations of both species in low proportions (Table 4). Most of them possessed an individual proportion of membership between 0.80 and 0.95. A relatively high number of individuals with 0.80<P(X|K)<0.95 were observed in the pure population of *Q. coccifera* in Kannaviou. Individuals with proportions of membership lower than 0.80 were rarer. Ten *Q. alnifolia* individuals within an overall total of 351, and eight *Q. coccifera* individuals out of 209, displayed scores of less than 0.80. They also occurred in most populations, whether pure or sympatric. Finally, two designated *Q. alnifolia* trees from the mixed stand were assigned to the cluster of *Q. coccifera* with low proportions of membership (< 0.8; Table 4). We noticed that genotypic admixture was partly due to the occurrence of one species' diagnostic alleles in the other species. This was more obvious among admixed *Q. alnifolia* individuals probably because this species was more uniform at most loci included into the Bayesian analysis of genetic structure.

*cpDNA population diversity*

A total of 12 chlorotypes were observed. These were assigned numbers from 1 to 12 (Table 5). Seven out of twelve chlorotypes were shared between the two species, whereas no species-specific lineages could be observed (Figure 4). Some chlorotypes were frequent in both *Q. alnifolia* and *Q. coccifera*. In particular, chlorotypes 2 and 4 were observed in both species with frequencies varying from 14.7 to 21.6 %. Chlorotypes 3, 10 and 11 occurred only in *Q. alnifolia* and 5, 8 and 12 only in *Q. coccifera*. However, their frequency was mostly low (< 10 %) with one exception – chlorotype 10 in *Q. alnifolia* (16.0 %). The geographic distribution of all variants is illustrated in Figure 5. Some chlorotypes dominate in a specific region, like chlorotype 2 that occurs in the eastern and chlorotype 4 that occurs in the western populations of both *Q. alnifolia* and *Q. coccifera* (Figure 5). Others exhibit a rather local distribution, for example chlorotype 9, which is scattered amongst three *Q. coccifera* and one *Q. alnifolia* population. In the mixed population, cpDNA sharing is limited. Only chlorotype 7 occurs in both species (frequent occurrence in *Q. alnifolia* and rare in *Q. coccifera*), whereas chlorotypes 5 and 12 are confined to *Q. coccifera* and 6 to *Q. alnifolia*, respectively (Figure 5).

Further spatial analysis of geographic patterns provided insights regarding the spatial genetic structure of *Q. alnifolia* and *Q. coccifera*. In general, significant spatial genetic barriers do not coincide with species boundaries. For instance, no significant differentiation among one of the northernmost *Q. coccifera* populations (c6) and two eastern populations of *Q. alnifolia* (a5, a7) was observed (Figure 6). As shown in Figure 5, these populations consist of the same chlorotypes although they belong to different species. Moreover, these populations grow on two different mountain ranges (a5 and a7 on Troodos; c6 on Pentadaktylos) separated by the central plain where oak species are currently absent. Chlorotype 2 is shown to dominate these populations, whereas chlorotype 1 is present at a much lower frequency. Genetic barriers were named using the letters A-G and were sorted by decreasing genetic differentiation in terms of $R_{ST}$ (Figure 6). All barriers were significant, as supported by 96-100 bootstraps. The highest differentiation was observed between population c1, which was separated by the genetic barrier A from its neighbouring populations (c4, a1 and a3). Barrier B was extensive and divided the remaining populations into western (c3, c4, a1, a3 and a4) and eastern parts (c5-7, a5-7). The remaining barriers (C-G) were local. For example, barrier C separated the population a5 from the population c5.

DISCUSSION

Contrasting differentiation patterns were revealed by the analysis of nuclear and chloroplast microsatellite markers in natural populations of *Q. alnifolia* and *Q. coccifera* throughout Cyprus. A high degree of interspecific differentiation was detected by means of nuclear SSRs. Species discrimination was highly significant both in pure populations and in sympatry. On the other hand, cpDNA haplotypes were shared between species and demonstrated partly shared geographic structures. These results are comparable to other reports of nuclear and cpDNA diversity among related oak species, such as in section *Robur* including *Q. petraea*, *Q. robur*, *Q. pubescens* and other species in Europe (Petit et al. 2002; Curtu et al. 2007b), *Q. ilex* and *Q. suber* in the Western Mediterranean (López de Heredia et al. 2005; Soto et al. 2007; Mir et al. 2009), or *Q. douglasii* and *Q. lobata* in North America (Craft and Ashley 2006).

All nuclear microsatellite loci analyzed in our study were polymorphic. Genetic diversity varied among loci, as well as between species. Two loci, QrZAG11 and 112 were highly differentiating between species. Due to the prevalence of one allele in *Q.*

*alnifolia* versus a diverse pattern in *Q. coccifera*, locus QrZAG11 showed high levels of differentiation between the two species. Locus QrZAG112 was highly discriminant due to the predominance of different diagnostic alleles with high frequencies in each species. Similar patterns of differentiation have been reported between several interfertile pairs of oaks, involving different marker loci in each case. For instance, five '$F_{ST}$ outlier' SSR loci have been found to distinguish *Q. petraea* from *Q. robur* (Scotti-Saintagne et al. 2004). Results from Soto et al. (2007) support that locus QpZAG9 is likewise highly discriminant between *Q. ilex* and *Q. suber*.

High differentiation of nuclear genomes between *Q. alnifolia* and *Q. coccifera* in our study was also supported by multilocus analyses. Species were the main component of genetic variation in AMOVA, whereas subdivision within species accounted for a minimal part of the total variation. Structure analysis further supported the high differentiation between the two species at the population level. Whether pure or in sympatry, populations of *Q. alnifolia* and *Q. coccifera* displayed high proportions of membership to their own species cluster. Individuals with admixed genotypes occurred in low proportions in both mixed and pure stands. This result indicates that interspecific hybridization is limited. Higher introgression rates in contact zones cannot be supported. In such cases the sympatric population should have been more admixed and would have included a higher proportion of admixed individuals in comparison to the pure populations. Increased genetic introgression in sympatry has been observed in mixed populations of closely related oak species. For instance, in the case of *Q. petraea* and *Q. pubescens*, hybridization leads to an increased occurrence of introgressed genotypes in sympatry (Curtu et al. 2007b, Gugerli et al. 2008, Salvini et al. 2009).

On the other hand, multilocus genetic differentiation of nuclear SSRs within species was found to be low, but significant. Moreover, both $F_{ST}$ values and levels of significance were found to be higher in *Q. coccifera*. In contrast to *Q. alnifolia* which is confined to the volcanic rock formations of the Troodos Mountains, *Q. coccifera* grows on several different geological substrates. The pure population of *Q. coccifera*, where the large sample was taken from, grows on sedimentary rock formations. The mixed stand and the pure population of *Q. alnifolia* both occur on intrusive rocks (diabase) of the ophiolite sequence of Troodos. The distinctive soil conditions between the two *Q. coccifera* stands may account for the more pronounced intraspecific differentiation found in this species. Similar results were reported from Lorenzo et al. 2009, where high genetic differentiation was found among populations of *Q. suber* occurring under differential edaphic conditions on the island of Minorca. Furthermore, *Q. alnifolia* was found to be generally less diverse at the studied loci, which can be attributed to its highly restricted habitat. Similar findings based on isoenzyme markers were presented by Toumi and Lumaret (2001). Also in the paradigm of *Q. ilex* and *Q. suber*, the latter species, confined to acidic soils, is less diverse than the former, which has greater edaphic amplitude (Burgarella et al. 2009).

Anthropogenic influences may have played a significant role in the formation of genetic structures within and between *Q. alnifolia* and *Q. coccifera*. The habitat of *Q. alnifolia* is restricted to the rugged landscape of the Troodos Mountains. There is no evidence of a major disturbance due to human activities since the initial agropastoral settlement of Cyprus during the Neolithic in these areas (about 10800 BP; Butzer and Harris 2007). On the contrary, the fertile soils where *Q. coccifera* is primarily distributed were intensively exploited by humans, either for agriculture or for grazing. For instance, large areas in the region southwest of Troodos, where the sample of the large *Q. coccifera* population was taken from, have been intensively used since antiquity for grape cultivation leading to a high fragmentation of the indigenous plant societies (Christou 2001). This fragmentation may have led to a higher degree of interpopulation genetic differentiation within this species, due to demographic events.

Regarding cpDNA, maternal lineages are shared between *Q. alnifolia* and *Q. coccifera*. Haplotypic diversity of these species in Cyprus is high compared to related species growing in the Mediterranean Basin (see Fineschi et al. 2005; López de Heredia et al. 2005 and Magri et al. 2007 for an overview about *Q. coccifera*, *Q. ilex* and *Q. suber* in the western Mediterranean). Geographic and genetic barriers do not coincide and the distribution of chlorotypes is rather locally structured. These structures may partially reflect landscape barriers. Such barriers between the western, the central and the eastern portions of the Troodos Mountains may have contributed to the *Q. alnifolia* structures. In the case of *Q. coccifera*, cpDNA data support connectivity of the two western populations (in Akamas (c3) and Lysos (c4)) which might have been remnants of a denser shrubland prior to fragmentation due to human activities in this region. The complex distribution of the chlorotypes may reflect different waves of colonization from the past. In the case of western Mediterranean sclerophyllous oaks, distribution of cpDNA lineages still reflect the westward colonization which took place during the tertiary (Lumaret et al. 2002, López de Heredia et al. 2007). In these species, local structures are often shared among species, although hybridization levels are low (Belahbib et al. 2001, Burgarella et al. 2009, Mir et al. 2009). Furthermore, the scattered occurrence of some chlorotypes throughout the island may reflect some sporadic long distance dispersal events. Given that *Q. coccifera*, in contrast to *Q. alnifolia*, has been used as a pasture and for the production of crimson dye, human seed transfer cannot be ruled out. This could be the case in the large population whose prevalent chlorotype (chlorotype 9) is not common for the south-western region of Cyprus.

The high degree of cpDNA sharing between *Q. alnifolia* and *Q. coccifera* and the occurrence of some common regional chlorotypes (e.g. chlorotype 2 in central and north-eastern Cyprus and chlorotypes 4 and 6 in the west) support the hypothesis that the two species have exchanged their genetic variants through gene flow. However, a restricted cpDNA sharing in the mixed stand from our study does not support a high degree of current introgressive hybridization, which would balance the chlorotype frequencies between the species. This hypothesis agrees with the low introgression of nuclear genomes in this stand, which is not higher than in pure stands. A similar situation in an oak community with four interfertile species was reported by Curtu et al. (2007b), who observed that *Quercus robur* showed the lowest degree of nuclear admixture compared to *Q. frainetto*, *Q. petraea* and *Q. pubescens*. Furthermore, its predominant chlorotype was not found in any individual of the other species. In our study, the predominant chlorotype of *Q. coccifera* in the mixed stand (chlorotype 12) was not found in any one of the analyzed *Q. alnifolia*.

Nevertheless, once hybridization between two related species is possible, there is the potential of succeeding introgression. In the present study, we provided evidence that initial hybridization between *Q. alnifolia* and *Q. coccifera* occurs in nature. The four designated hybrids possessed admixed genotypes and their chlorotypes were observed in both *Q. coccifera* (chlorotype 5) and *Q. alnifolia* (chlorotype 7; results not shown) from the same stand. Moreover, hybrids produced large amounts of acorns (field observations). Occasional hybridization and backcrossings might have happened several times in the past leading to local exchanges of chlorotypes. The imprints of such historic introgression events may have remained intact since the Pleistocene glaciation and did not lead to severe extinctions in the Mediterranean Basin (Jimenez et al. 2004, López de Heredia et al. 2007). The low dispersal ability of cpDNA might account for the prevalence of specific chlorotypes in a given geographic region and occasional introgressive hybridization may have led to exchange between the two species. Despite being distantly related, *Q. ilex* and *Q. suber* were able to exchange their cpDNA, although hybridization between them is not common (Belahbib et al. 2001, Mir et al. 2009). Introgression of their nuclear and chloroplast genomes has been shown to be independent (Mir et al. 2009). Besides selection, reproductive barriers due to speciation also limit genetic introgression (Belahbib *et al.* 2001; Boavida et al. 2001). Speciation of *Q. alnifolia* and *Q. coccifera* were already distinct during Miocene (Palamarev 1989). Separate evolution for a long time may have led to the occurrence of reproductive barriers limiting hybridization.

ACKNOWLEDGMENTS

We would like to thank Andreas Christou, Constantinos Kounnamas, Salih Gücel, Zacharias Triftarides, Savvas Protopapas, Petros Anastasiou and the Department of Forests, Republic of Cyprus personnel, for their assistance during plant collections. We are grateful to Dr. Unai López de Heredia for stimulating discussions. This research was conducted in partial fulfilment for the doctorate degree of the first author at Albert-Ludwigs University of Freiburg. During that time the first author was supported by doctorate scholarships from DAAD and the State of Baden-Württemberg. Partial financial assistance to Filippos Aravanopoulos in the form of two cooperative grants of the Ministry of Natural Resources of Cyprus and the Aristotle University of Thessaloniki is gratefully acknowledged.

TABLES

**Table 1 – Diversity measures within populations of *Q. alnifolia* and *Q. coccifera***

Sample size (N), allelic richness ($R_{65}$ ; rarefaction size= 65), observed ($H_0$) and expected ($H_e$) heterozygosity, inbreeding coefficients ($F_{IS}$; * P≤0.05,** P≤0.01, *** P≤0.001) and estimated frequency of null alleles. Only the two largest populations of each species are presented. Null allele frequencies are given only if they are significant.

| Popu-lation | Locus | *Quercus alnifolia* | | | | | | *Quercus coccifera* | | | | | |
|---|---|---|---|---|---|---|---|---|---|---|---|---|---|
| | | N | $R_{65}$ | Ho | He | $F_{IS}$ | null | N | $R_{65}$ | Ho | He | $F_{IS}$ | null |
| Pure | QpZAG9 | 95 | 9.853 | 0.653 | 0.686 | 0.054 | | 96 | 17.047 | 0.833 | 0.841 | 0.014 | |
| | QrZAG11 | 91 | 3.972 | 0.209 | 0.220 | 0.055 | | 92 | 12.486 | 0.761 | 0.795 | 0.049 | |
| | QrZAG96 | 95 | 3.000 | 0.484 | 0.487 | 0.011 | | 96 | 5.564 | 0.354 | 0.341 | -0.032 | |
| | QrZAG112 | 95 | 4.585 | 0.432 | 0.442 | 0.030 | | 96 | 4.964 | 0.375 | 0.349 | -0.070 | |
| | QpZAG119 | 76 | 22.477 | 0.368 | 0.924 | 0.605 *** | 0.297 | 92 | 22.829 | 0.804 | 0.897 | 0.109 *** | 0.055 |
| | **Multilocus** | | | 0.429 | 0.552 | 0.228 | | | | 0.626 | 0.645 | 0.035 | |
| Mixed | QpZAG9 | 207 | 8.898 | 0.681 | 0.720 | 0.057 | | 66 | 11.985 | 0.848 | 0.851 | 0.011 | |
| | QrZAG11 | 206 | 4.929 | 0.150 | 0.156 | 0.040 | | 65 | 13.000 | 0.646 | 0.816 | 0.216 ** | 0.104 |
| | QrZAG96 | 207 | 3.844 | 0.527 | 0.519 | -0.013 | | 66 | 4.985 | 0.379 | 0.409 | 0.082 | |
| | QrZAG112 | 207 | 5.785 | 0.386 | 0.423 | 0.088 | | 66 | 4.000 | 0.091 | 0.116 | 0.225 | |
| | QpZAG119 | 188 | 27.272 | 0.420 | 0.949 | 0.559 *** | 0.277 | 65 | 17.000 | 0.615 | 0.831 | 0.267 *** | 0.134 |
| | **Multilocus** | | | 0.433 | 0.553 | 0.220 | | | | 0.516 | 0.605 | 0.154 | |

**Table 2 – AMOVA summary.**

Analysis of Molecular Variance (AMOVA) based on nuclear DNA data from the four large populations (a1, a2, c1 and c2). Variation was dissected to the following components: species, population within species, within population and within individuals (overall). Significance was tested by a permutation procedure (* P≤0.05,** P≤0.01, *** P≤0.001).

| Source of variation | Degrees of freedom | Sum of squares | Variance components (significance) | Fixation indices | Percentage of variation |
|---|---|---|---|---|---|
| **Among species groups** | 1 | 211.145 | 0,49320*** | 0.333 ($F_{CT}$) | 33.26 |
| **Among populations, within species** | 2 | 5.463 | 0,00823** | 0.008 ($F_{SC}$) | 0.55 |
| **Within populations** | 461 | 465.227 | 0,02770* | 0.028 ($F_{IS}$) | 1.87 |
| **Within individuals (overall)** | 465 | 443.500 | 0,95376*** | 0,357 ($F_{IT}$) | 64.32 |

**Table 3 – Pairwise population comparisons**

Pairwise $F_{ST}$s between the four largest populations are presented (a1, a2, c1 and c2). Values from each locus are shown separately and a multilocus value is given as well. Significance was tested by permuting individuals between the compared populations (* P≤0.05, ** P≤0.01, *** P≤0.001).

| Populations compared | INTRASPECIFIC | | INTERSPECIFIC | | | |
|---|---|---|---|---|---|---|
| | a1-a2 | c1-c2 | a1-c1 | a1-c2 | a2-c1 | a2-c2 |
| Locus | | | | | | |
| QpZAG9 | 0.0063(n.s.) | 0.0076* | 0.1465*** | 0.1350*** | 0.1404*** | 0.1186*** |
| QrZAG11 | 0.0184** | 0.0118* | 0.3285*** | 0.3488*** | 0.4324*** | 0.4657*** |
| QrZAG96 | 0.0015(n.s.) | 0.0014(n.s.) | 0.0670*** | 0.0606*** | 0.0747*** | 0.0637*** |
| QrZAG112 | 0.0028(n.s.) | 0.0497*** | 0.5729*** | 0.6888*** | 0.5724*** | 0.6612*** |
| Overall | 0.0049* | 0.0127*** | 0.3104*** | 0.3504*** | 0.3299*** | 0.3642*** |

**Table 4 – Proportions of membership to species group and number of admixed individuals**

Results based on the Bayesian analysis with the Structure software are presented. Population localities, species and number of sampled individuals per population are given in the first three columns. Under P(X|K) proportions of membership of each population to each one of the two inferred clusters are given. A threshold of P(X|K)= 0.95 was set to distinguish 'pure' from admixed individuals. Admixed genotypes are separated into those with individual membership proportions (P(X|K)) between 0.8 and 0.95 symbolized with N(0.8<P<0.95) and those with respective values between 0.5 and 0.8 symbolized with N(0.5<P<0.8).

| | | | *Quercus alnifolia* cluster | | | | | | *Quercus coccifera* cluster | |
|---|---|---|---|---|---|---|---|---|---|---|
| | | | Group membership | Number of individuals per class of membership proportion | | | | | | Group membership |
| Population | Sp. | N | P(X|K) | N (P>0.95) | N (0.8<P<0.95) | N (0.5<P<0.8) | N (0.5<P<0.8) | N (0.8<P<0.95) | N (P>0.95) | P(X|K) |
| Stavros | | 96 | 0,975 | 90 | 4 | 2 | 0 | 0 | 0 | 0,025 |
| Kionia | | 10 | 0,954 | 8 | 1 | 1 | 0 | 0 | 0 | 0,046 |
| Papoutsa | | 10 | 0,952 | 8 | 1 | 1 | 0 | 0 | 0 | 0,048 |
| Madari | *Q. alnifolia* | 10 | 0,947 | 7 | 2 | 1 | 0 | 0 | 0 | 0,053 |
| Karvounas | | 10 | 0,986 | 10 | 0 | 0 | 0 | 0 | 0 | 0,014 |
| Kykkos | | 8 | 0,974 | 7 | 1 | 0 | 0 | 0 | 0 | 0,026 |
| Kambos | | 207 | 0,966 | 182 | 18 | 5 | 2 | 0 | 0 | 0,034 |
| Kambos (hybrids) | | 4 | 0,270 | 0 | 0 | 1 | 1 | 2 | 0 | 0,730 |
| Kambos | | 66 | 0,022 | 0 | 0 | 0 | 1 | 2 | 63 | 0,978 |
| Pentadaktylos | | 12 | 0,015 | 0 | 0 | 0 | 0 | 0 | 12 | 0,986 |
| Lapithos | | 10 | 0,041 | 0 | 0 | 0 | 1 | 0 | 9 | 0,959 |
| Akamas | *Q. coccifera* | 10 | 0,052 | 0 | 0 | 0 | 1 | 1 | 8 | 0,948 |
| Lysos | | 10 | 0,060 | 0 | 0 | 0 | 1 | 2 | 7 | 0,940 |
| Kalopanayiotis | | 5 | 0,034 | 0 | 0 | 0 | 1 | 1 | 4 | 0,966 |
| Kannaviou | | 96 | 0,033 | 0 | 0 | 0 | 3 | 12 | 81 | 0,967 |

**Table 5 – Multilocus chloroplast haplotypes and their frequencies within each species**

Allele sizes for each locus are presented in columns. Frequencies of each chlorotype within species are given at the bottom of the table.

|  |  | Chlorotypes | | | | | | | | | | | |
|---|---|---|---|---|---|---|---|---|---|---|---|---|---|
|  | Locus | 1 | 2 | 3 | 4 | 5 | 6 | 7 | 8 | 9 | 10 | 11 | 12 |
| Allele sizes | ccmp2 | 235 | 235 | 234 | 234 | 234 | 234 | 234 | 235 | 235 | 234 | 234 | 234 |
|  | µdt1 | 81 | 81 | 81 | 81 | 81 | 81 | 81 | 82 | 82 | 82 | 82 | 82 |
|  | µdt3 | 119 | 119 | 120 | 120 | 120 | 120 | 120 | 119 | 119 | 120 | 120 | 120 |
|  | µcd4 | 95 | 95 | 93 | 94 | 93 | 94 | 95 | 93 | 93 | 94 | 94 | 93 |
|  | µcd7 | 83 | 84 | 83 | 83 | 84 | 84 | 84 | 83 | 84 | 83 | 84 | 84 |
|  | µkk3 | 95 | 95 | 95 | 95 | 95 | 95 | 95 | 94 | 94 | 95 | 95 | 95 |
| Chlorotype frequency | *Q.alnifolia* | 10% | 27% | 3% | 20% | 0% | 13% | 1% | 0% | 6% | 16% | 4% | 0% |
|  | *Q.coccifera* | 10% | 20% | 0% | 19% | 3% | 4% | 1% | 9% | 27% | 0% | 0% | 7% |

**FIGURES**

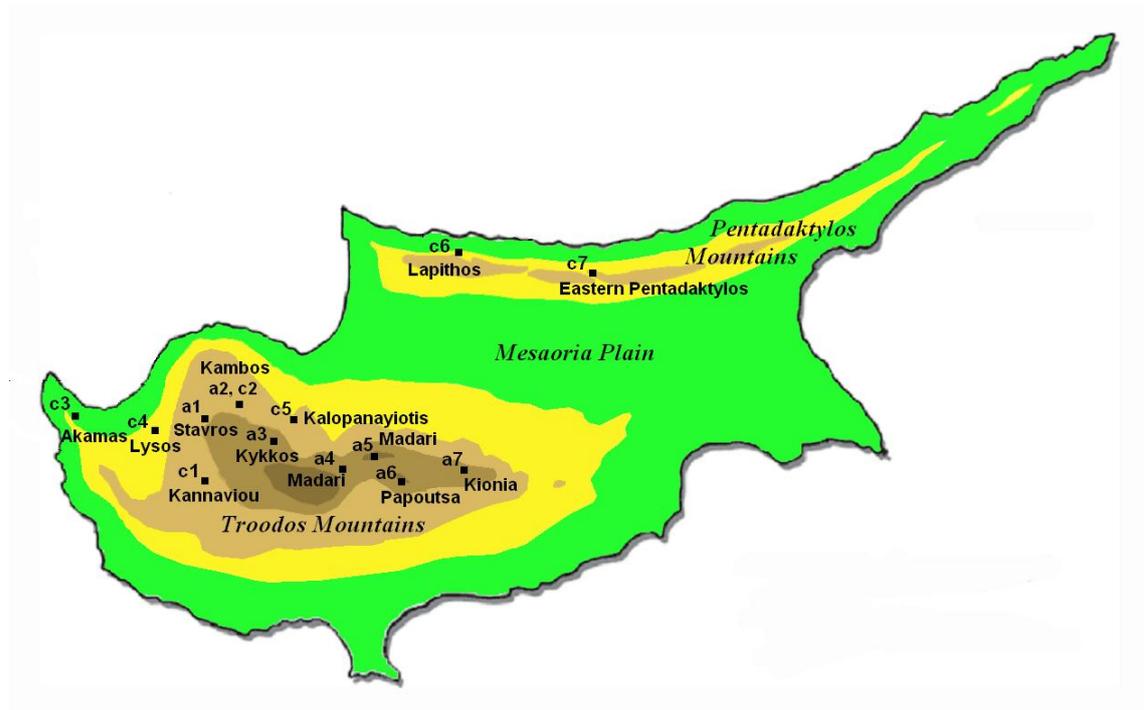

Fig. 1 – Geophysical map of Cyprus showing locations of all sampled populations. Stands are marked as follows: *Quercus alnifolia*: a1-7, *Q. coccifera*: c1-7. Locality names corresponding to each population are given.

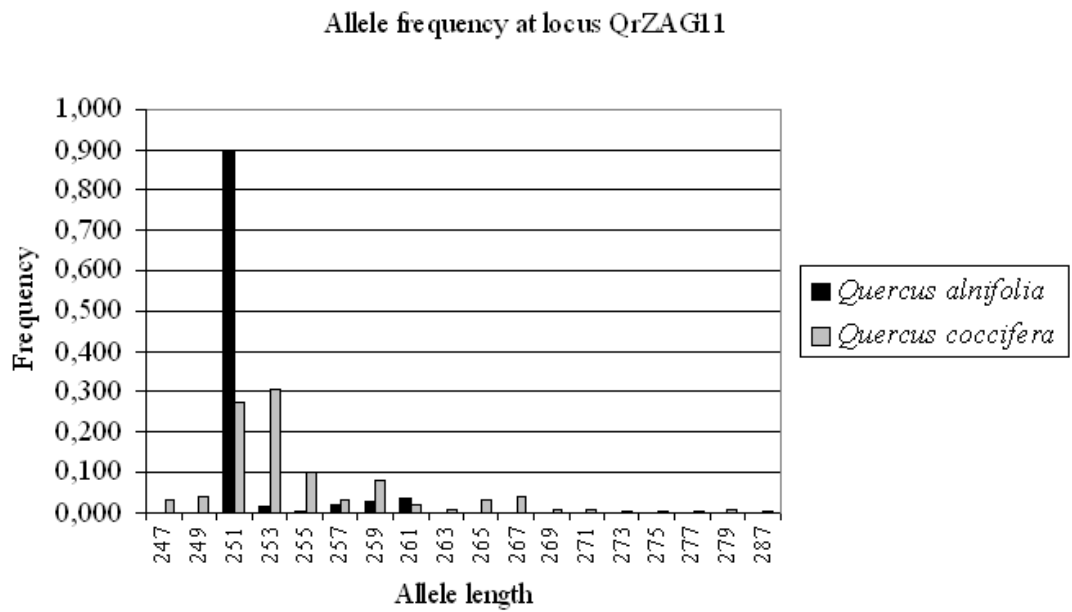

Fig. 2 – Allele frequencies at loci QrZAG11 in *Q. alnifolia* and *Q. coccifera*. *Q. alnifolia* is represented by black filled bars and *Q. coccifera* by grey filled bars.

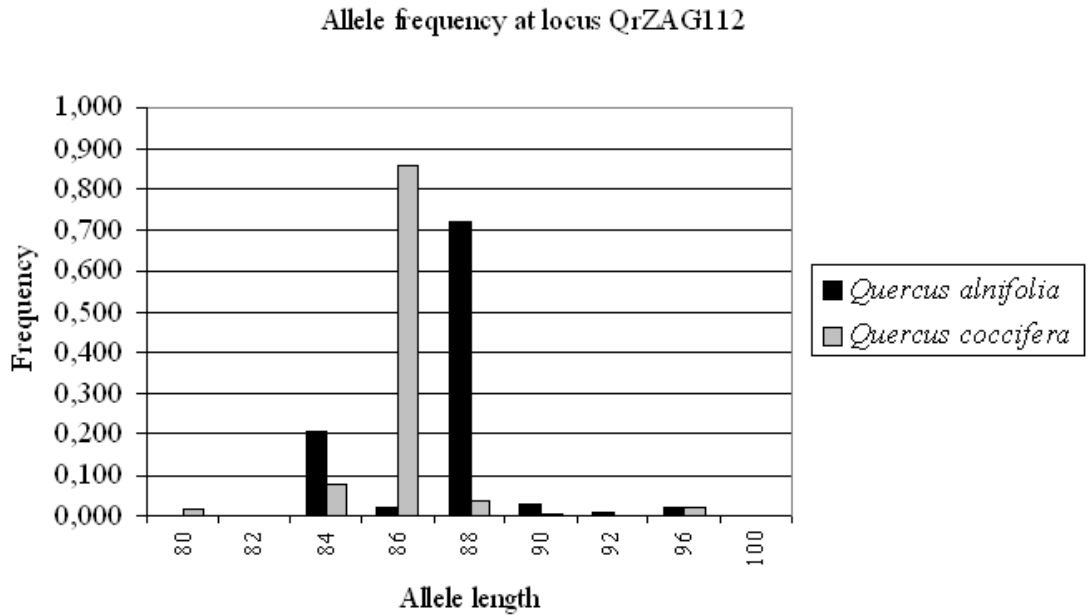

Fig. 3 – Allele frequencies at locus QrZAG112 in *Q. alnifolia* and I. *Q. alnifolia* is represented by black filled bars and *Q. coccifera* by grey filled bars.

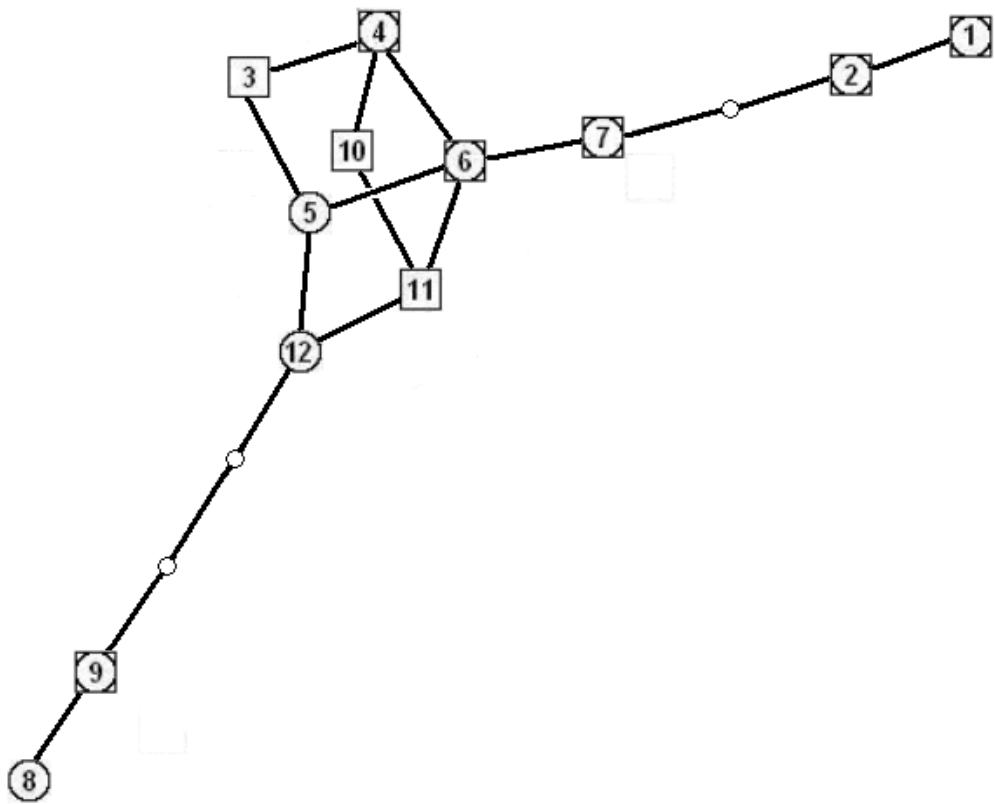

Fig. 4 – Minimum spanning network of haplotypes. Haplotype nomenclature follows Table 5. Chlorotypes occurring in *Q. alnifolia* are marked with a square and chlorotypes observed in *Q. coccifera* are designated by a circle. Small circles represent missing or unsampled chlorotypes.

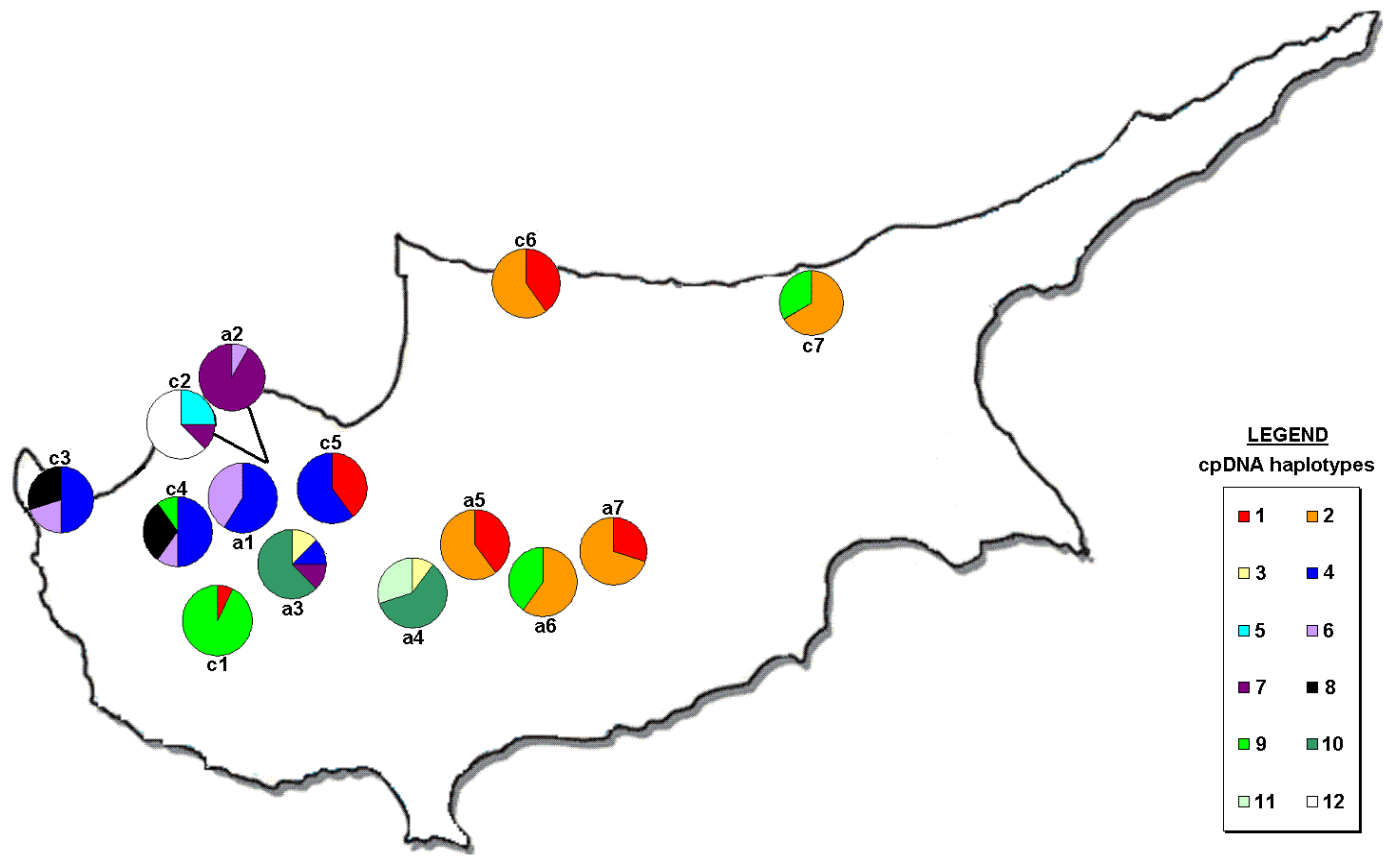

Fig. 5 – Frequency of chlorotypes across sampled stands in Cyprus. Each chlorotype is represented by one colour and its frequency is illustrated in a pie diagram for each population. Stands are marked as follows: Quercus alnifolia: a1-7, Q. coccifera: c1-7

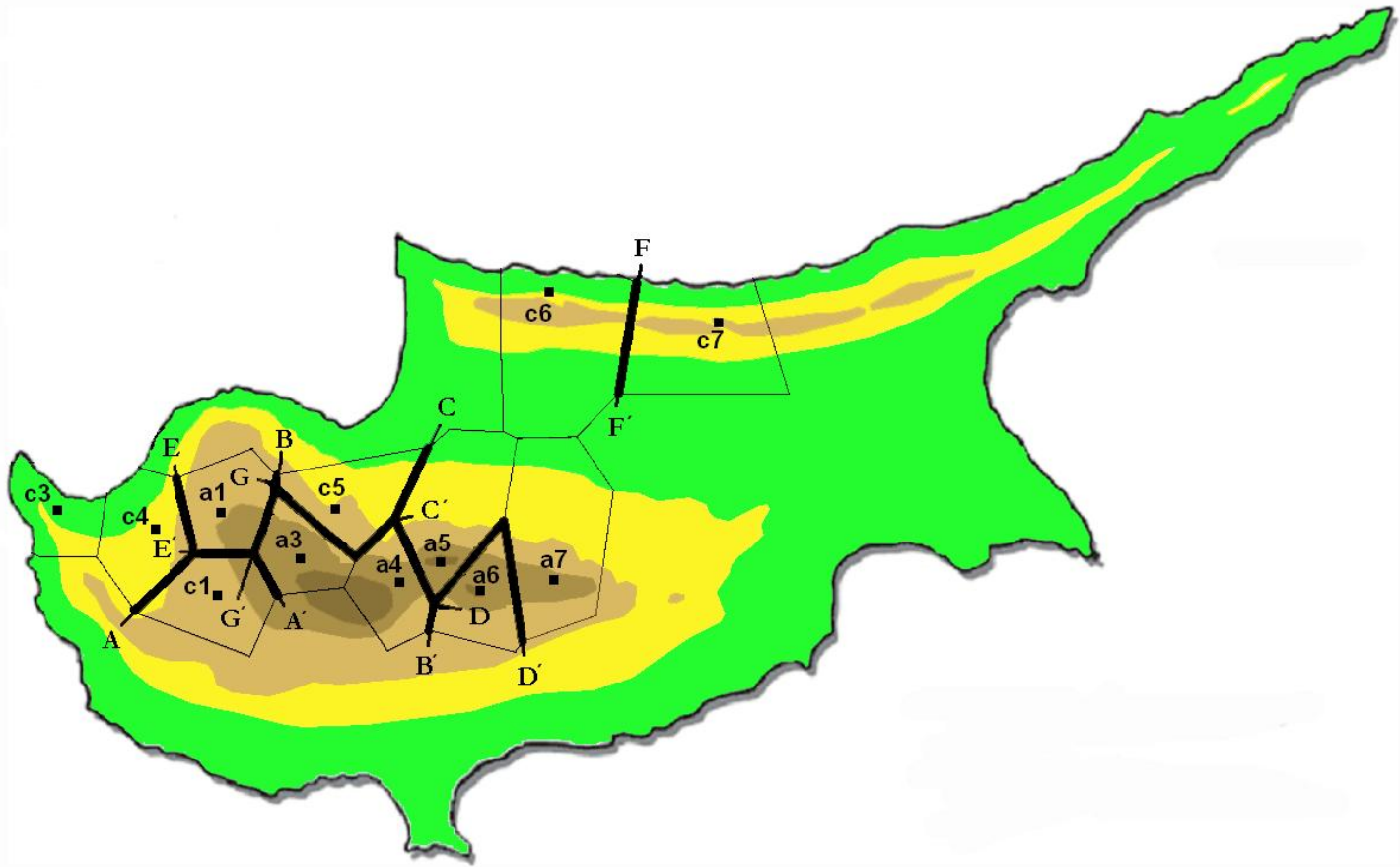

Fig. 6 – Spatial genetic barriers between populations using $R_{ST}$ based on cpDNA variation. Polygons corresponding to each population were placed in order to present genetic barriers among them. Seven barriers are shown with thick black lines at the edges of each polygon. The beginning of a barrier is marked by the letters A-G and thy end by A'-G'.